\documentclass[aps,twocolumn,numerical]{revtex4-2}
\usepackage[T1]{fontenc}
\usepackage[utf8]{inputenc}
\usepackage{color}
\usepackage{amsmath}
\usepackage{graphicx}
\usepackage[unicode=true,pdfusetitle,
 bookmarks=true,bookmarksnumbered=false,bookmarksopen=false,
 breaklinks=false,pdfborder={0 0 0},pdfborderstyle={},backref=false,colorlinks=true]
 {hyperref}
\hypersetup{
 linkcolor=blue,citecolor=blue}
\begin{document}
\title{Triply degenerate nodal line and tunable contracted-drumhead surface
state\\ in a tight-binding model}
\author{Yi-Ru Wang}
\author{Gui-Bin Liu}
\email[Corresponding author: ]{gbliu@bit.edu.cn}

\address{Centre for Quantum Physics, Key Laboratory of Advanced Optoelectronic
Quantum Architecture and Measurement (MOE), School of Physics, Beijing
Institute of Technology, Beijing 100081, China}
\address{Beijing Key Laboratory of Nanophotonics and Ultrafine Optoelectronic
Systems, School of Physics, Beijing Institute of Technology, Beijing
100081, China}
\begin{abstract}
The study of topological semimetals has been extended to more general
topological nodal systems such as metamaterials and artificial periodic
structures. Among various nodal structures, triply degenerate nodal
line (TNL) is rare and hence lack of attention. In this work, we have
proposed a simple tight-binding model which hosts a topological non-trivial
TNL. This TNL not only has the drumhead surface states as usual nodal
line systems, but also has surface states which form a contracted-drumhead
shape. And the shape and area of this contracted-drumhead can be tuned
by the hopping parameters of the model. This provides an effective
way to modulate surface states as well as their density of states,
which can be important in future applications of topological nodal
systems. 
\end{abstract}
\maketitle

\section{Introduction}

In recent years, topological semimetals have become a frontier topic
in condensed matter physics because of their promising applications
in electronics, spintronics, and optics \citep{r2,r3,r4,r30,r31,r32,r33,r34,r1}.
According to the dimensions of the degenerate manifolds in $k$-space
formed by band crossings, topological semimetals are divided into
nodal point semimetals such as Weyl \citep{r1,r5,Xu_Hasan_2015_349_613__Discovery,Xu_Hasan_2015_11_748__Discovery},
Dirac \citep{Wang_Fang_2012_85_195320__Dirac,Wang_Fang_2013_88_125427__Three,Liu_Chen_2014_343_864__Discovery,Liu_Chen_2014_13_677__stable,Neupane_Hasan_2014_5_3786__Observation},
or triple point semimetals \citep{Lv_Ding_2017_546_627__Observation,Ma_Ding_2018_14_349__Three},
nodal line semimetals \citep{Fang_Fu_2015_92_81201__Topological,Huang_Duan_2016_93_201114__Topological,Xu_Weng_2017_95_45136__Topological,r9},
and nodal surface semimetals \citep{Wu_Yang_2018_97_115125__Nodal,Fu_Ding_2019_5_6459__Dirac,Yang_Zhang_2019_10_5185__Observation,Chen_Duan_2020_20_5400__Nodal}.
Due to the non-trivial topological band structure, Weyl (Dirac) semimetals
can exhibit Fermi arc surface states \citep{r1,r5,r31} connecting
different Weyl-node (Dirac-node) projections on a two-dimensional
(2D) surface Brillouin zone (BZ), and nodal line semimetals can exhibit
another special surface state --- drumhead surface state (DSS) \citep{r17,r29,Bian_Hasan_2016_93_121113__Drumhead,Huang_Duan_2016_93_201114__Topological,Xu_Weng_2017_95_45136__Topological,Belopolski_Hasan_2019_365_1278__Discovery}
on a 2D surface BZ. In fact, these non-trivial topological properties
are not limited to being present in semimetals, because they originate
from the nodal band structures and they also exist in other systems
such as metals \citep{Xu_Hasan_2015_347_294__Observation,Wang_Bernevig_2016_117_56805__MoTe$_2$,Zhu_Soluyanov_2016_6_31003__Triple,Sun_Chang_2017_96_45121__Coexistence},
optical crystals \citep{Gao_Zhang_2018_9_950__Experimental,Yang_Zhang_2018_359_1013__Ideal,Wang_Yao_2016_93_61801__Topological},
phononic crystals \citep{Li_Liu_2017_14_30__Weyl,Zhang_Fang_2018_120_16401__Double},
mechanical systems \citep{r36}, and circuit systems \citep{r18,r23,Luo_Weng_2018_2018_6793752__Topological}.

For topological nodal line materials, doubly degenerate Weyl nodal
line \citep{r8,r9,r10,r11} and quadruply degenerate Dirac nodal line
\citep{r12,r13,r14,r15} have been broadly studied, and the DSS has
been observed in both these two kinds of materials. However, about
triply degenerate nodal line (TNL), there is very little research
on it. The only such research we can find is Ref. \citep{r39} by
Liu \textit{et al.} in 2021. In Ref. \citep{r39}, two TNL models
were proposed, one of which is nontopological and the other is topological
according to the existence of Fermi arc topological surface states.
However, no DSS was reported for the TNL models in Ref. \citep{r39}.
Accordingly, in this work we aim to construct a tight-binding (TB)
model with TNL and investigate its DSS.

However, it's almost impossible to construct a TNL model based on
real crystalline materials, because real crystalline materials are
constrained by the symmetries of (magnetic) space groups and systematic
studies on the possible emergent particles from band crossings have
shown that no TNL exists under various (magnetic) space groups \citep{Yu_Yao_2022_67_375__Encyclopedia,Liu_Yao_2022_105_85117__Systematic,Zhang_Yao_2022_105_104426__Encyclopedia}.
Then, to construct a TNL model one has to get rid of the constraints
by (magnetic) space groups. This can be achieved in artificial systems
such as metamaterials, circuit systems, and mechanical systems, because
when described by TB models the effective hoppings in these systems
can be tuned at will, e.g. adjusting the connection mode among circuit
components or changing the coupling strength through springs \citep{r23,r35,r36}.

In this work, we first constructed a three-band TB model hosting TNL
by designing the hoppings. Then, we calculated the Berry phase and
Zak phase to check the topological non-triviality of the TNL. Surface
states on two different surfaces, i.e. $(010)$ and $(\bar{1}10)$,
were studied via both semi-infinite systems and slab models. The usual
DSS was found on the (010) surface. However, on the $(\bar{1}10)$
surface we noticed a new kind of DSS, whose drumhead is not a complete
one but a contracted one. The tuning of this DSS with contracted drumhead
was also studied by varying the hopping parameters of the model.

\section{Model and Method}

\begin{figure}[tb]
\centering{}\includegraphics[width=8cm]{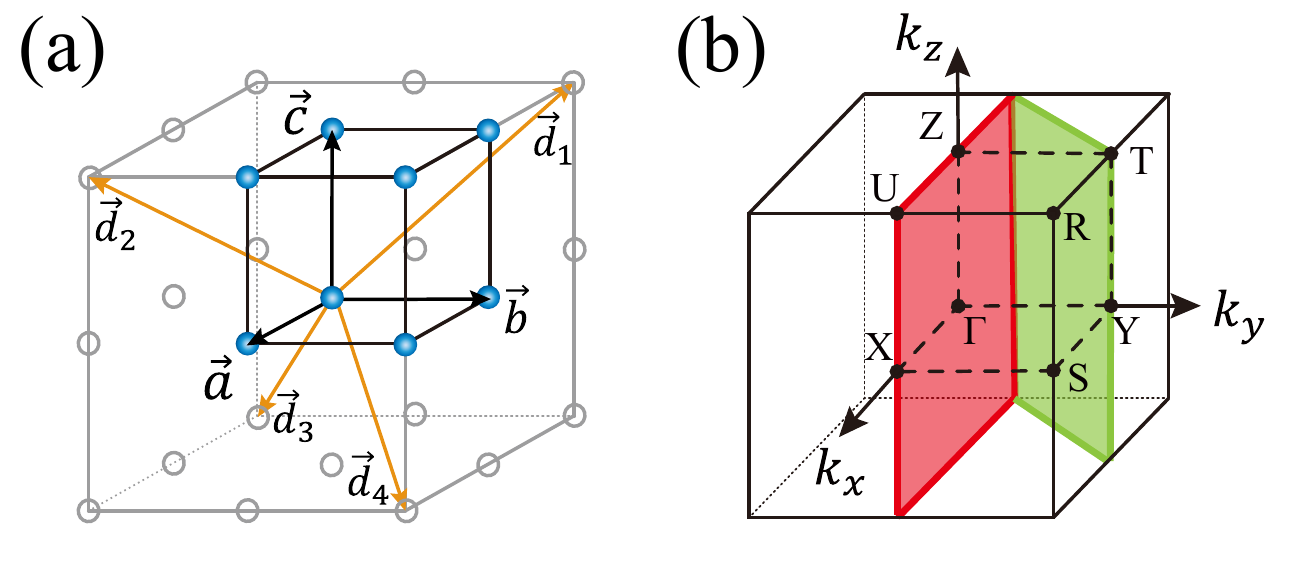} \caption{(a) Unit cell (black) and hopping vectors of the model. (b) The bulk
BZ and its projections to $(010)$ surface (red) and $(\bar{1}10)$
surface (green).}
\label{Fig:1}
\end{figure}

The model is constructed based on a simple cubic lattice whose basis
vectors $\vec{a}$, $\vec{b}$, and $\vec{c}$ are equal in magnitude
and along $x$, $y$, and $z$ direction respectively, as shown in
Fig.~\ref{Fig:1}(a). Only one atom with three orbitals (here called
$\phi_{1}$, $\phi_{2}$, $\phi_{3}$) is considered in each cell.
With the hopping between orbitals $\phi_{i}(\vec{r})$ and $\phi_{j}(\vec{r}-\vec{d})$
denoted as $h_{ij}(\vec{d})$, we choose the following hoppings for
the model.
\begin{equation}
h_{11}(0)=t_{0},\ h_{22}(0)=-t_{0},\ h_{33}(0)=2t_{0}\label{eq:onsite}
\end{equation}
\begin{equation}
h_{11}(\pm\alpha)=\frac{t_{1}}{2},\ h_{22}(\pm\alpha)=-\frac{t_{1}}{2},\ h_{33}(\pm\alpha)=t_{1}\label{eq:nn}
\end{equation}
\begin{equation}
h_{23}(\pm\vec{b})=\pm\frac{t_{2}}{2},\ \ h_{12}(\pm\vec{d}_{i})=\pm\frac{t_{3}}{8}\ (i=1,2,3,4)\label{eq:tnn}
\end{equation}
in which $\alpha=\vec{a},\vec{b},\vec{c}$. Then the Hamiltonian of
the TB model is
\begin{equation}
\begin{aligned}H=(t_{0}+t_{1}\cos k_{x}+t_{1}\cos k_{y}+t_{1}\cos k_{z})\lambda_{1}\ \ \\
+t_{2}\sin k_{y}\lambda_{2}+t_{3}\sin k_{x}\sin k_{y}\sin k_{z}\lambda_{3}
\end{aligned}
\label{eq:H}
\end{equation}
in which $\lambda_{1}$, $\lambda_{2}$, and $\lambda_{3}$ are the
following three matrices respectively: 
\begin{equation}
\begin{aligned}\left(\begin{array}{ccc}
1 & 0 & 0\\
0 & -1 & 0\\
0 & 0 & 2
\end{array}\right),\ \ \left(\begin{array}{ccc}
0 & 0 & 0\\
0 & 0 & i\\
0 & -i & 0
\end{array}\right),\ \ \left(\begin{array}{ccc}
0 & i & 0\\
-i & 0 & 0\\
0 & 0 & 0
\end{array}\right)\end{aligned}
.\label{eq:lambda}
\end{equation}
One can easily see that, when $k_{y}=0$, Eq. (\ref{eq:H}) becomes
a diagonal matrix whose diagonal elements can be null simultaneously.
This implies that a TNL can exist in the plane $k_{y}=0$ under suitable
values of $t_{0}$ and $t_{1}$. The key feature of the hoppings that
results in this TNL is that $h_{ii}(0)/h_{ii}(\pm\alpha)$ keeps constant
for $i=1,2,3$. This is a special request that cannot be derived from
symmetries of (magnetic) space group.

The surface density of state (SDOS) was obtained by calculating the
surface Green's function of semi-infinite system using the Wanniertools
package \citep{r25}. The input data for Wanniertools was prepared
by the MagneticTB package \citep{r24}. To investigate the surface
states, we also constructed TB slab models of 80 layers using the
PythTB package \citep{r26}. In order to judge whether a state is
a surface state, we first define the topmost five layers on each side,
A or B, of the slab model as ``surface layers'', and then define
the following quantity $\eta$ to characterize the degree to which
a state is a surface state 
\begin{equation}
\begin{aligned}\eta=\begin{cases}
[w_{{\rm A}}+w_{{\rm B}}-\frac{1}{8}]/\frac{7}{8}, & w_{{\rm A}}+w_{{\rm B}}>\frac{1}{8}\\
0, & w_{{\rm A}}+w_{{\rm B}}\le\frac{1}{8}
\end{cases}\end{aligned}
\label{eq:eta}
\end{equation}
where $w_{{\rm A/B}}$ represents the wavefunction weight within the
surface layers of side A/B of the slab. For a bulk state wavefunction,
it is periodic and its weights are equally distributed within all
the 80 layers, in which case $w_{{\rm A}}=w_{{\rm B}}=5/80$ and $\eta=0$.
For a perfect surface state, wavefunction is totally localized within
the surface layers, leading to $w_{{\rm A}}=w_{{\rm B}}=1/2$ and
$\eta=1$. By means of $\eta$, a state can be determined as a strong
(or typical) surface state if its $\eta$ is greater than a critical
value $\eta_{{\rm c}}$ and in this paper $\eta_{{\rm c}}=0.5$ is
adopted.

\section{Results}

\begin{figure}[t]
\centering{}\includegraphics[width=8cm]{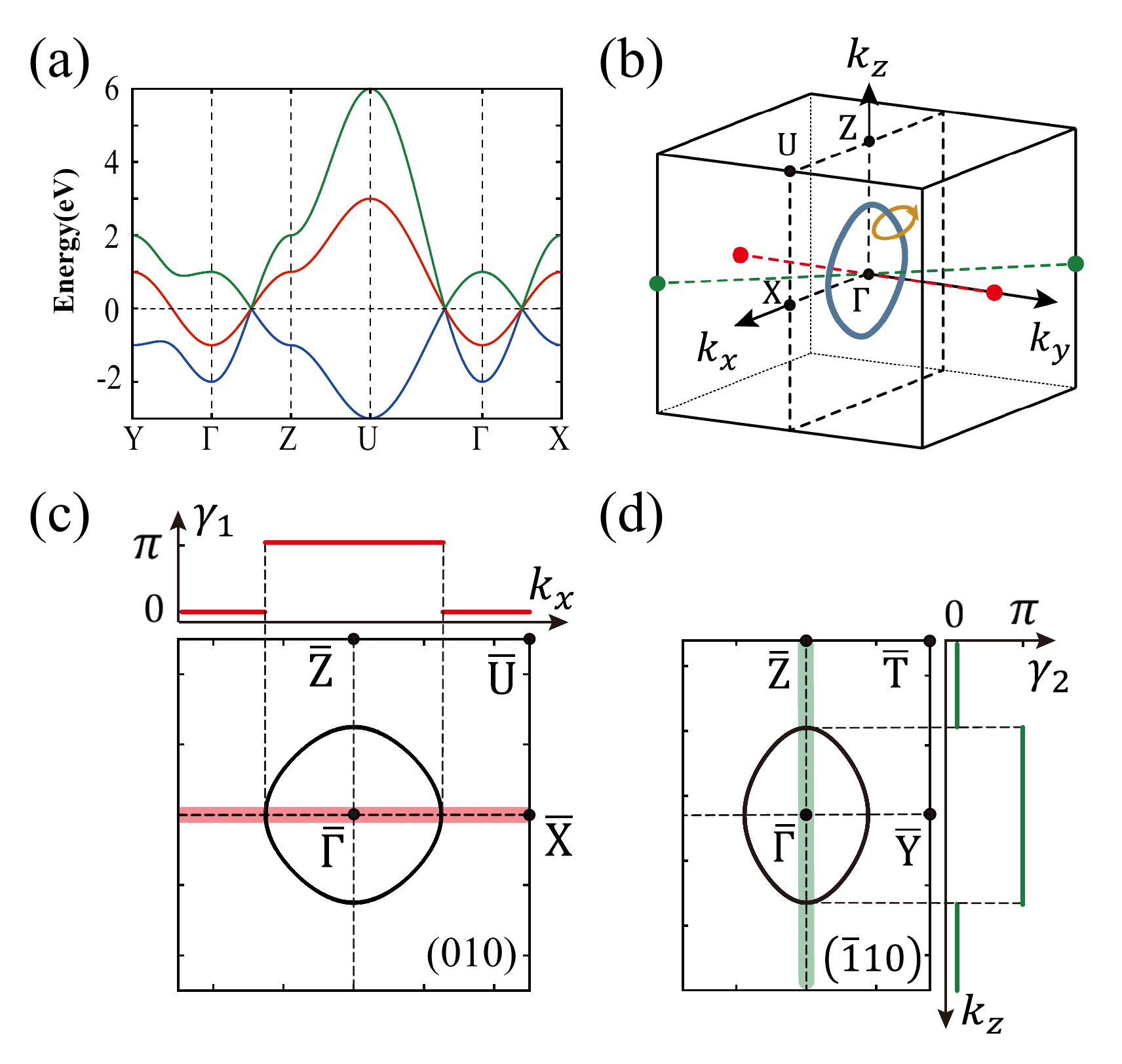}\caption{(a) The bulk band structure of the model. (b) The TNL (thick blue
nodal ring) and the \textit{k}-point path (small orange loop) for
calculating the Berry phase. The red (green) dashed line is the integral
path for Zak phase $\gamma_{1}$ at $k_{x}=0$ ($\gamma_{2}$ at $k_{z}=0)$.
(c,d) The TNL projections onto (c) $(010)$ and (d) $(\bar{1}10)$
surface BZs, and the Zak phases (c) $\gamma_{1}(k_{x})$ and (d) $\gamma_{2}(k_{z})$.}
\label{Fig:2}
\end{figure}

If not otherwise stated, the parameters $t_{0}=2$, $t_{1}=-1$, and
$t_{2}=t_{3}=1$ are used for the model and the unit is eV for all
energies. The bulk energy bands are shown in Fig. \ref{Fig:2}(a)
in which the $k$ points are defined in Fig. \ref{Fig:1}(b). In this
model, the TNL is actually an approximately circular nodal ring in
the $k_{y}=0$ plane, as shown in Fig. \ref{Fig:2}(b). To check the
topological properties of the TNL, we first calculated the Berry phase
defined on a closed \textit{k}-point loop enclosing the TNL, with
fully gapped energies, as shown by the small orange loop in Fig.~\ref{Fig:2}(b).
The Berry phase is calculated using the Wilson loop approach \citep{r27}
and the result is $\pi$, which shows the topological non-triviality
of the TNL.

\begin{figure*}[t]
\centering{}\includegraphics[width=12cm]{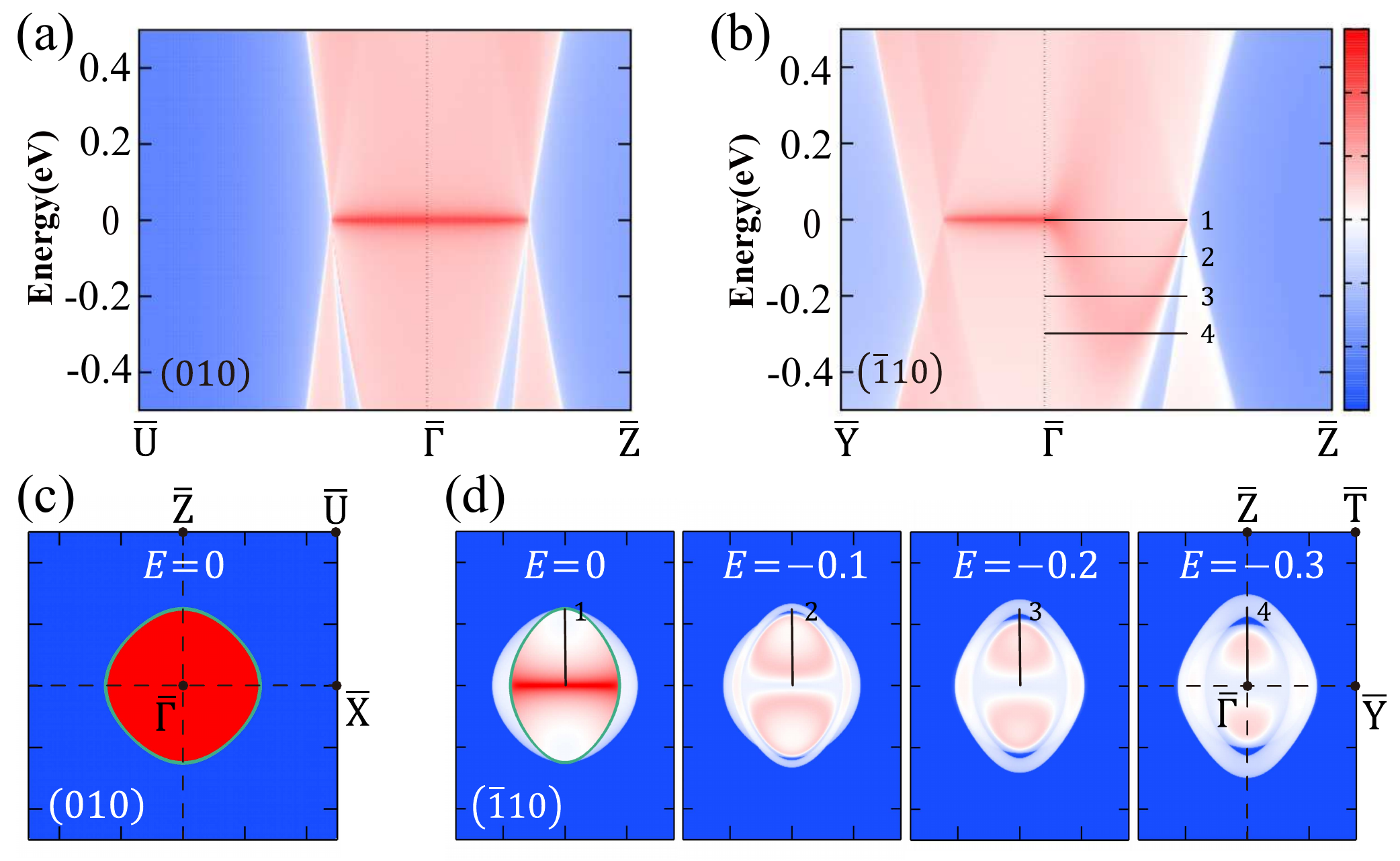}\caption{Topological surface states given by SDOS for the semi-infinite systems
terminated with (a,c) (010) and (b,d) $(\bar{1}10)$ surfaces. (a,b)
Continuous energy resolved SDOS. (c) The constant energy slice at
$E=0$ for the $(010)$ surface system. (d) The constant energy slices
at $E=0,-0.1,-0.2,-0.3$ for the $(\bar{1}10)$ surface system, in
which the cutting lines 1--4 correspond to the ones in (b). The green
lines in (c,d) are the projections of the TNL.}
\label{Fig:3}
\end{figure*}

\begin{figure*}[t]
\centering{}\includegraphics[width=16cm]{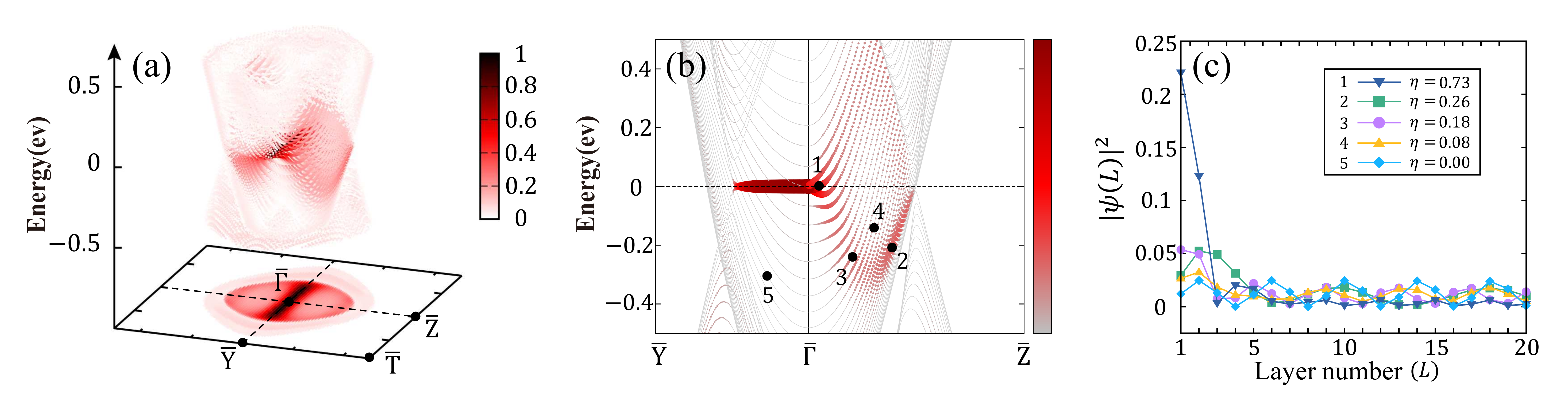}\caption{The $(\text{\ensuremath{\bar{1}}}10)$ slab model. (a) The distribution
of $\eta$ (the degree of surface state) represented by color for
all states with energy in the range $[-0.5,0.5]$. The largest $\eta$
at each $k$-point is also projected onto the surface BZ. (b) Energy
bands with $\eta$ represented by both color and the point size. (c)
The squared wavefunctions with respect to the layer number (only 20
out of the total 80 layers are shown) for the five states marked in
(b).}
\label{Fig:4}
\end{figure*}

Besides, we also calculated the Zak phase \citep{r28,Xiao_Niu_2010_82_1959__Berry},
which is the Berry phase defined in a one-dimensional BZ along a certain
direction. Two Zak phases are investigated here. The first one $\gamma_{1}(k_{x})$
is defined along the line from $(k_{x},-\frac{1}{2},0)$ to $(k_{x},\frac{1}{2},0)$
with the $k_{x}=0$ case shown by the red dashed line in Fig.~\ref{Fig:2}(b),
in which the $k$-point coordinates are in unit of $2\pi/a$. The
second one $\gamma_{2}(k_{z})$ is defined along the line from $(\frac{1}{2},-\frac{1}{2},k_{z})$
to $(-\frac{1}{2},\frac{1}{2},k_{z})$ with the $k_{z}=0$ case shown
by the green dashed line in Fig.~\ref{Fig:2}(b). The calculated
Zak phase $\gamma_{1}(k_{x})$ ($\gamma_{2}(k_{z})$) is shown in
the top (right) panel of Fig. \ref{Fig:2}(c) (Fig. \ref{Fig:2}(d))
whose $k_{x}$ ($k_{z}$) axis corresponds to the red (green) thick
line in the 2D projective BZ of (010) (($\bar{1}$10)) surface shown
in the corresponding bottom (left) panel. We can see that, for both
$\gamma_{1}$ and $\gamma_{2}$, the non-trivial $\pi$ Zak phase
emerges only when the integral path of Zak phase traverses the nodal
ring, i.e. the TNL here; otherwise the Zak phase is zero. According
to the bulk-edge correspondence \citep{Rhim_Bardarson_2017_95_35421__Bulk},
this change of topological properties from inside to outside the nodal
ring implies the existence of topological surface states inside the
projected nodal ring on both (010) and $(\bar{1}10)$ surfaces.

To explore the topological surface states of a certain surface, the
semi-infinite system terminated with that surface should be constructed.
Two surfaces (010) and $(\bar{1}10)$ are studied here, where the
(010) surface is parallel to the nodal ring but the ($\bar{1}$10)
surface is not. Fig. \ref{Fig:3}(a) shows the SDOS of the (010) surface
system, whose surface states all have a constant energy (zero) and
form a flat drumhead shape. This typical DSS is clearly demonstrated
by the SDOS with constant energy slice at $E=0$, as shown in Fig.~\ref{Fig:3}(c).
The green ring in Fig.~\ref{Fig:3}(c) represents the front projection
of the TNL, and its interior is full of surface states. We call this
kind of DSS as ``full DSS''. From the result that both the Zak phases
$\gamma_{1}$ and $\gamma_{2}$ equal $\pi$ inside the TNL projections,
one may expect that full DSS also exists in the $(\bar{1}10)$ surface
system. However, the SDOS of the $(\bar{1}10)$ surface system shown
in Fig. \ref{Fig:3}(b) and (d) demonstrates results different from
the expectation. Especially, the leftmost panel of Fig. \ref{Fig:3}(d)
shows ``contracted-drumhead surface state (CDSS)'', in which the
surface states do not fill completely the interior of the TNL oblique
projection (the green ellipse). The weak surface state feature at
other energies as shown in other panels of Fig. \ref{Fig:3}(d) also
supports this result.

To further explore the CDSS in the $(\bar{1}10)$ surface system,
a slab model of 80 layers terminated with $(\bar{1}10)$ surface is
studied and its energy bands are shown in Fig.~\ref{Fig:4}(b), in
which the degree of surface state $\eta$ defined in Eq. (\ref{eq:eta})
is also shown by both the point size and color for each state. In
addition, Fig.~\ref{Fig:4}(a) shows the the distribution of $\eta$
for all states within the energy range $[-0.5,0.5]$ in the whole
surface BZ and the projection of $\eta$ onto the surface BZ. Fig.~\ref{Fig:4}(b)
corresponds to Fig.~\ref{Fig:3}(b), but here we can access the wavefunction
of any state of the slab model. The wavefunctions of the five states
marked in Fig.~\ref{Fig:4}(b) have descending $\eta$ from 0.73
to 0 and the distributions of their weights with respect to layer
number are given in Fig.~\ref{Fig:4}(c). We can see that the state
1 with $\eta=0.73$ is a strong surface state with most wavefunction
localized within the surface layers. At the other extreme, state 5
with $\eta=0$ distributes periodically and hence is a bulk state.
As for the states 2--4, they have nonzero but small $\eta$. Although
they contain surface state components or may be called weak surface
states, they are more like bulk states. From Figs.~\ref{Fig:4}(a)
and \ref{Fig:4}(b) we can see that strong surface states exist only
near zero energy. Consequently, even if the projection of $\eta$
in Fig.~\ref{Fig:4}(a) selects the largest $\eta$ for each $k$-point
within the energy range $[-0.5,0.5]$, it is not much different from
the case considering only zero energy and it exhibits similar shape
to the $E=0$ panel in Fig.~\ref{Fig:3}(d).

\begin{figure}[t]
\centering{}\includegraphics[width=8cm]{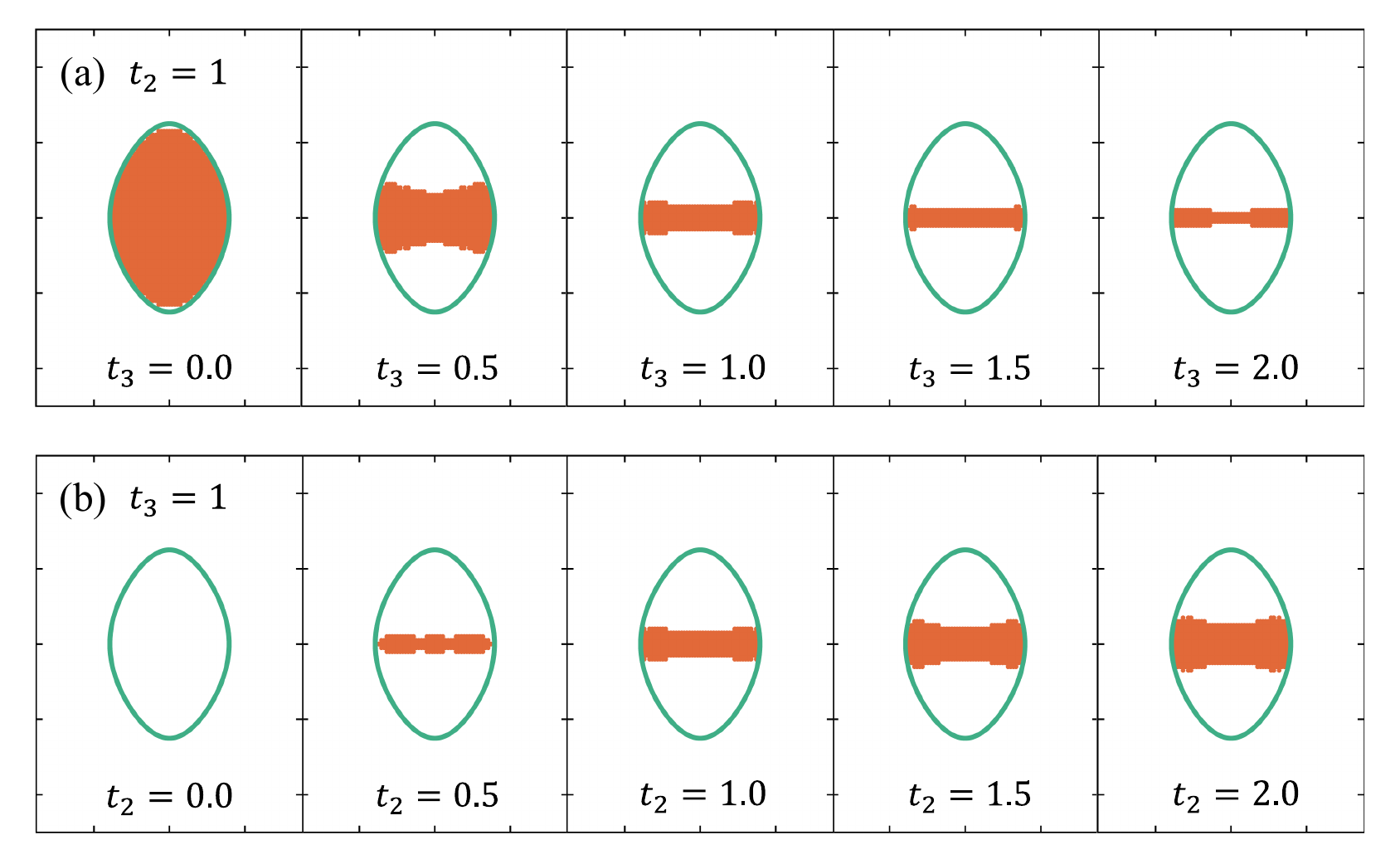}\caption{The $k$-point distribution (orange area) of the CDSS for the $(\bar{1}10)$
slab model with different parameters. (a) $t_{3}=0,\,0.5,\,1.0,\,1.5,\,2.0$
with $t_{2}=1.0.$ (b) $t_{2}=0,\,0.5,\,1.0,\,1.5,\,2.0$ with $t_{3}=1.0.$
The green ellipse denotes the projection of the TNL. Other two parameters
are $t_{0}=2$ and $t_{1}=-1$ for all cases.}
\label{Fig:5}
\end{figure}

Because the weak surface states are much like bulk states, they are
not efficient in most applications which require large SDOS. Thus,
only strong surface states need to be considered, and the shape of
the CDSS can be revealed by the distribution of the $k$-points at
which strong surface states exist. Fig.~\ref{Fig:5} shows the distribution
of the $k$-points of strong surface states, i.e. the shape of CDSS,
under different model parameters. We can see that the shape of CDSS
for the $(\bar{1}10)$ surface can be tuned by the hoppings $t_{2}$
and $t_{3}$ efficiently, namely, increasing $t_{3}$ makes the surface
states changing from a full DSS to a CDSS with smaller and smaller
areas (Fig.~\ref{Fig:5}(a)), and inversely, increasing $t_{2}$
will increasing the area of CDSS from zero (Fig.~\ref{Fig:5}(b)).
This provides an effective route to tune the SDOS and the shape of
surface state in topological nodal systems.

\section{Conclusions}

In summary, we have proposed a simple TB model hosting TNL and studied
its topological properties. Both Berry phase and Zak phase demonstrate
that the TNL is topological non-trivial. This TNL model not only has
a full DSS as usual topological nodal line systems, but also has a
CDSS which we first noticed. The CDSS exists on the $(\bar{1}10)$
surface and its area can be tuned efficiently by both the model parameters
$t_{2}$ and $t_{3}$. Our model demonstrates an effective way to
tune the amount and density of states of the surface states, which
will expand the potential applications of topological nodal line systems.

\section{ACKNOWLEDGMENTS}

This work is supported by the National Natural Science Foundation
of China with Grant Nos. 12274028, 52161135108, and 12234003 and the
National Key R\&D Program of China with Grant No. 2022YFA1402603.

\end{document}